\documentclass[twocolumn,aps,prc]{revtex4}
\usepackage{epsfig}

\newcommand{\beq}{\begin{eqnarray}}
\newcommand{\eeq}{\end{eqnarray}}

\begin{document}

\title{Resonant states of neutron-rich $\Lambda$ hypernucleus $^7_{\Lambda}$He}

\author{E. Hiyama}
\author{M. Isaka}
\affiliation{Nishina Center for Accelerator-Based Science,
Institute for Physical and Chemical Research (RIKEN), Wako 351-0198, Japan}
\author{M. Kamimura}
\affiliation{Department of Physics, Kyushu University, Fukuoka, 812-8581, Japan}
\affiliation{Nishina Center for Accelerator-Based Science,
Institute for Physical and Chemical Research (RIKEN), Wako 351-0198, Japan}
\author{T. Myo}
\affiliation{General Education, Faculty of Engineering, Osaka Institute of 
Technology, Osaka, 535-8585, Japan}
\author{T. Motoba}
\affiliation{Laboratory of Physics, Osaka Electro-Communication University, 
Neyagawa 572-8530, Japan}
\affiliation{Yukawa Institute for Theoretical Physics,
Kyoto University, Kyoto 606-8317, Japan}

%

\begin{abstract}
The structure of  neutron-rich $\Lambda$ hypernucleus, $^7_{\Lambda}$He
is studied within the framework of an $\alpha +\Lambda +n+n$ four-body cluster 
model. We predict second $3/2^+$ and $5/2^+$ states, corresponding to a
$0s$ $\Lambda$ coupled to the second $2^+$ state of $^6$He, as narrow resonant 
states with widths $\Gamma \sim 1$ MeV to be at 0.03 MeV and 0.07 MeV 
respect to the $\alpha +\Lambda +n+n$ threshold. From an estimation of the
differential cross section for the $^7{\rm Li} (\gamma,K^+) ^7_{\Lambda}$He
reaction, there is a possibility to observe these state at JLab in the future.
We also calculate the second $2^+$ state of $^6$He as resonant state
within the framework of an $\alpha +n+n$ three-body cluster model.
Our result is $2.81$ MeV with $\Gamma =$4.63 MeV with respect to the 
$\alpha +n+n$ threshold. This energy position is $\sim 1$ MeV
higher, and with a much broader decay width, than the recent SPIRAL data.
It is suggested that an experiment at JLab to search for the second $3/2^+$
and $5/2^+$ states of $^7_{\Lambda}$He would provide an opportunity
to confirm the second $2^+$ state of the core nucleus $^6$He.
\end{abstract}

\maketitle

\section{Introduction}

In  2013,  a neutron rich $\Lambda$ hypernucleus, $^7_{\Lambda}$He
was observed via the ($e,e'K^+)$ reaction, and an observed $\Lambda$ 
separation energy of $B_{\Lambda}= 5.68\pm 0.03 ({\rm stat.})\pm
0.25({\rm sys.})$ MeV was reported~\cite{JLAB}.
This observation stimulated us to study neutron-rich
$\Lambda$ hypernuclei because in light nuclei near the neutron drip line, 
interesting phenomena concerning neutron halos have been observed. When a 
$\Lambda$ particle is added to such nuclei, it is expected that the 
resultant hypernuclei will become more stable against neutron
decays due to the attraction of $\Lambda N$ interaction and the fact
that there is no Pauli exclusion effect between nucleons and a $\Lambda$ particle.
This phenomenon is one of the 'gluelike' roles of $\Lambda$ particle.

Before this measurement, we recently predicted in Refs.~\cite{Hiyama96,Hiyama2009}
the energy spectra of $^7_{\Lambda}$He in the bound-state region
within the frameworks of a $^5_{\Lambda}{\rm He}+n+n$ three-body model and 
an $\alpha +\Lambda+n+n$ four-body cluster model.
The core nucleus $^6$He is known to be a typical neutron halo nucleus:
the two-neutron separation energy is 0.975 MeV.
The $\Lambda$ participation in the bound state of such
a halo nucleus results in a more stable ground state of the hypernucleus.
We predicted that the binding energy of the ground state is 5.36 MeV
within the $\alpha +\Lambda +n+n$ four-body model which is consistent with the
recent data within the error bar.
For this ground state, we have another interesting insight related to
the charge-symmetry breaking (CSB) components in the $\Lambda N$ interaction.
It is considered that the most reliable evidence for CSB appears in the
$\Lambda$-separation energies ($B_{\Lambda}$) of the $A=4$ hypernuclei
with $T=1/2$ ($^4_{\Lambda}$H and $^4_{\Lambda}$He). Then,
the CSB effects are attributed to the separation-energy difference
$\Delta _{\rm CSB}= B_{\Lambda}(^4_{\Lambda}$He)$-B_{\Lambda}(^4_{\Lambda}$H),
the experimental values of which are $0.35 \pm 0.06$ MeV and
$0.24 \pm 0.06$ MeV for the
ground ($0^+$) and excited $(1^+$) states, respectively.
It is also likely that CSB contribution affects to the
binding energy of $^7_{\Lambda}$He
and the experimental research at JLab on the $^7_{\Lambda}$He
was motivated by this question.
 Since consistent understanding of the CSB in the $\Lambda N$
 interaction has not yet been obtained, further experimental studies of
$^4_{\Lambda}$H will be carried out in near future at JLab and Mainz,
and $^4_{\Lambda}$He at J-PARC.

The experimental study of   $^7_{\Lambda}$He was performed again
in 2009 with 5 times more statistics (JLab E05-115 experiment).
The preliminary result gives more accurate binding energy of
the ground state, and the first excited state ($3/2^+_1$  or $5/2^+_1$)
was observed for the first time,
which corresponds to the $2^+_1$ state of $^6$He core nucleus coupled
with $\Lambda (0s)$~\cite{Gogami}.
The observed $2^+_1$ state is located 0.827 MeV above the $\alpha +n+n$
breakup threshold with a decay width of $\Gamma =0.113$ MeV~\cite{Tilley}.
The coupling of a $\Lambda$ to this state leads to  bound states
of $^7_{\Lambda}$He. A high-resolution spectroscopy  experiment of
$\Lambda$ hypernuclei using an electron beam is one of the powerful
tools to produce bound and resonant  hypernuclear states.
Then, we have the following question:
`Is there a possibility to have other new hypernuclear states in
$^7_{\Lambda}$He?'
To answer this question,  it is necessary to look at the energy spectra
of $^6$He core nucleus before studying $^7_{\Lambda}$He.
The  observed data of $^6$He \cite{Tilley} reported  $0^+_1$
round state as a bound state and
the $2^+_1$ resonant state with $E_{\rm x}=1.797$ MeV, $\Gamma=0.113$ MeV.
To search the second $2^+$ state, some experiments were performed. For 
example, the charge-exchange reaction, $^6{\rm Li}(t,^3{\rm He})^6{\rm He}$, 
was studied to explore the excited states above the first 2+ 
state~\cite{Nakamura2000}. However, clear evidence of the second 2+ state was 
not obtained. In 2012, in Ref.~\cite{Mougeot}, the transfer reaction experiment
$p(^8{\rm He}, t) ^6$He shows an indication of the second $2^+$ state of $^6$He  
as a resonant state at $E_{\rm x}=2.6 \pm 0.3$ ($\Gamma=1.6 \pm 0.4$) MeV.

Theoretically, many authors studied energy spectra of $^6$He
with various theoretical approaches~\cite{Pieper,Hagen,Volya,Myo2011}.
Among them, for instance, one of the present authors (T.M.) and 
collaborators studied the energy spectra
of $^6$He within the framework of the cluster-orbital shell model
(COSM) for the $\alpha +n+n$ three-body system using the
complex scaling method (CSM), that is one of the powerful methods
to obtain resonant energies and decay widths accurately.
They reproduced the energies of the observed ground state
and the first $2^+$ state very well. Moreover, they obtained second $2^+$ state at
2.52 MeV above the $\alpha +n+n$ threshold with width $\Gamma=3.87$ MeV.

When a $\Lambda$ particle is added to such a resonant state,
due to a gluelike role of $\Lambda$ particle, it is likely to result
in narrower resonant states of $3/2^+_2$ and $5/2^+_2$ of $^7_{\Lambda}$He.
The prediction of energies of second $3/2^+$ and $5/2^+$ states
and decay widths would encourage further experimental investigation
of $^7_{\Lambda}$He at JLab. With this aim, we discuss 
resonant states for  $^6$He and $^7_{\Lambda}$He using CSM within
the framework of $\alpha +n+n$ and $\alpha +\Lambda +n+n$ three-
and four-body cluster models, respectively.

This article is organized as follows: In Sec II., the method and interactions 
used in the three-body and four-body calculation for
the $^6$He and $^7_{\Lambda}$H  are described. The numerical results and the
corresponding discussions are presented in Sec.  III.  A summary is given in
Sec. IV.

\section{Model and Interaction}

The models employed in this article are the same as those in our previous
work~\cite{Hiyama2009}.
We employ  the $\alpha +n+n$ three-body model for $^6$He
and the $\alpha +\Lambda +n+n$ four-body model
for $^7_{\Lambda}$He; as for the Jacobi-coordinate sets, see
Fig.~3 in Ref.~\cite{Hiyama96} and
Fig.~1 in Ref.~\cite{Hiyama2009}, respectively.

The Hamiltonian for $^6$He is written as
\begin{equation}
H = T + V_{N N} +
+ \sum_{i=1}^2 \big[  V_{\alpha N_i}
+ V^{\rm Pauli}_{\alpha N_i} \big]
\label{eq:Hamil-he6} \quad,
\end{equation}
and that for $^7_{\Lambda}$He is described by
\begin{equation}
H = T + V_{N N} + V_{\Lambda \alpha}
+ \sum_{i=1}^2 \big[ V_{\Lambda N_i} + V_{\alpha N_i}
+ V^{\rm Pauli}_{\alpha N_i} \big] \quad.
\label{eq:Hamil-7Li}
\end{equation}
Here, $T$ is the kinetic-energy operator and  $V^{\rm Pauli}_{\alpha N_i}$
stands for the Pauli principle between the two valence neutrons
and the neutrons in the $\alpha$ cluster.
The two-body $N$-$N$ ({\rm AV8}, T$=$1),
$\alpha$-$N$, $\alpha$-$\Lambda$ and $\Lambda$-$N$ interactions
are chosen so as to reproduce accurately the observed
properties of all the subsystems composed of
$NN (T$=$1)$, $\alpha N, \alpha \Lambda,  \alpha NN$ and $\alpha \Lambda N$;
the details are described in Ref.~\cite{Hiyama2009}.
Here, we do not include any charge-symmetry breaking interaction for
the $\Lambda N$ part. 

The total wave functions for $^6$He and $^7_{\Lambda}$He are
 described as a sum of amplitudes of the
rearrangement channels within the $LS$ coupling scheme, respectively:

\begin{eqnarray}
 \!\!\!\!\!\!\!\!   && \Psi_{JM}({^6{\rm He}})
       =  \sum_{c=1}^{3}
      \sum_{n,N}  \sum_{l,L}
       \sum_{S}
       C^{(c)}_{nlNL S I} \nonumber  \\
 \!\!\!\!\!\!\!\!     &  \times & {\cal A}_N
      \Big[
       \Phi(\alpha) 
     \big[ \chi_{\frac{1}{2}}(N_1)
       \chi_{\frac{1}{2}}(N_2)
       \big]_S  \nonumber \\
&\times &      \big[ \phi^{(c)}_{nl}({\bf r}_c)
         \psi^{(c)}_{NL}({\bf R}_c)\big]_I
          \Big]_{JM}.\;
\label{li6wavefunction}
\end{eqnarray}

\begin{eqnarray}
 \!\!\!\!\!\!\!\!   && \Psi_{JM}({^7_{\Lambda}{\rm He}})
       =  \sum_{c=1}^{9}
      \sum_{n,N,\nu}  \sum_{l,L,\lambda}
       \sum_{S,\Sigma,I,K}
       C^{(c)}_{nlNL\nu\lambda S\Sigma IK} \nonumber  \\
 \!\!\!\!\!\!\!\!     &  \times & {\cal A}_N
      \Big[
       \Phi(\alpha) \big[\chi_\frac{1}{2}(\Lambda)
     \big[ \chi_{\frac{1}{2}}(N_1)
       \chi_{\frac{1}{2}}(N_2)
       \big]_S \big]_\Sigma  \nonumber \\
&\times &      \big[ \big[ \phi^{(c)}_{nl}({\bf r}_c)
         \psi^{(c)}_{NL}({\bf R}_c)\big]_I
        \xi^{(c)}_{\nu\lambda} (\mbox{\boldmath $\rho$}_c)
        \big]_{K}  \Big]_{JM}.\;
\label{li7wavefunction}
\end{eqnarray}
Here the operator  ${\cal A}_N$
stands for antisymmetrization between the two valence neutrons.
$\chi_{\frac{1}{2}}(N_i)$
and $\chi_{\frac{1}{2}}(\Lambda)$ are the spin functions of
the $i$-th nucleon and $\Lambda$ particle.
$\Phi^{(\alpha)}$ is the wavefunction of the
$\alpha$ cluster having the $(0s)^4$ configuration.
As for the spatial basis functions $\phi_{nlm}({\bf r}_c)$, 
$\psi_{NLM}({\bf R}_c)$ and
$\xi^{(c)}_{\nu\lambda\mu} (\mbox{\boldmath $\rho$}_c)$,
we took the Gaussian basis functions
with the ranges in a geometric progression;
detailed is written in Refs.\cite{Hiyama2009,Hiyama2003}.

In this work, we focus on the resonant states of $^6$He and $^7_{\Lambda}$He.
We then employ the CSM~\cite{CSM-ref1,CSM-ref2,CSM-ref3,Ho,Moiseyev}.
The CSM and its application to nuclear physics  problems are extensively
reviewed in Refs.~\cite{Aoyama2006,myo2014} and references therein.
Using the CSM, one can directly obtain the energy $E_r$ and
the decay width $\Gamma$ of the $\alpha nn$ and $\alpha \Lambda nn$ systems
by solving the eigenvalue problem for the complex scaled Schr\"{o}dinger 
equation with a scaling angle $\theta$,
\begin{equation}
[H_{\alpha nn (\alpha \Lambda nn)}(\theta) -E(\theta)]
\Psi_{\alpha nn (\alpha \Lambda nn)}(\theta)=0,
\end{equation}
where the boundary condition of the many-body outgoing wave is
automatically satisfied for the resonance, giving
$E=E_r -i\Gamma/2$ that is, in principle, independent of $\theta$.
The complex scaled Hamiltonian $H_{\alpha nn (\alpha \Lambda nn)}(\theta)$
is obtained by making the radial coordinate transformation with
the common angle of $\theta$
\begin{equation}
   r_c \to r_c \,e^{i \theta}, \qquad
R_c \to R_c \,e^{i \theta}, \qquad
{\rho}_c \to    \rho_c \,e^{i \theta}
\end{equation}
in the Hamiltonian $H_{\alpha nn (\alpha \Lambda nn)}$ of the
$\alpha nn$ ($\alpha \Lambda nn$) system.
The non-resonant continuum states are obtained
on the $2\theta$-rotated line in the complex energy plane.

A great advantage of the CSM is that a resonant
state is described by an $L^2$-integrable wave function.
Therefore, the Gaussian basis functions mentioned above have been
successfully used in calculations of the
CSM~\cite{Aoyama2006}.


\section{Results and Discussion}

\begin{figure}[b]
\begin{center}
\epsfig{file=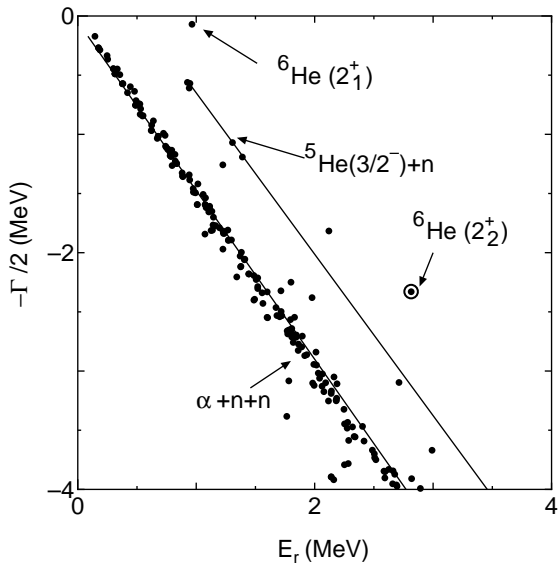,scale=0.45}
\end{center}
\caption{The $2^+_1$ and $2^+_2$ energy eigenvalue distributions of the
complex scaled Hamiltonian of $^6$He with $\theta=28^\circ$.
The energy is measured with respect to $\alpha +n+n$ threshold.
Two solid lines are $\alpha +n+n$ and $^5{\rm He}(3/2^-)+n$ threshold.}
\label{fig:scal-he6}
\end{figure}

\begin{figure}[b]
\begin{center}
\epsfig{file=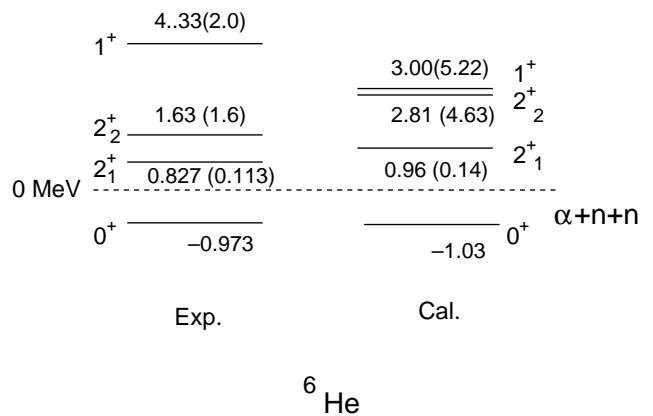,scale=0.45}
\end{center}
\caption{The calculated energy spectra of $^6$He
together with the experimental data. The energies in MeV
are measured with respect to the $\alpha +n+n$ threshold.
The values in parentheses are the decay widths $\Gamma$ in MeV. }
\label{fig:he6level}
\end{figure}

In Fig.~\ref{fig:scal-he6}, we show the distributions of
eigenenergies ($E_2^+ (\theta)=E_r-i\Gamma/2$)
of the complex-scaled Hamiltonian
$H_{\alpha n n}(\theta)$ for the $J^\pi=2^+$ states
of the $^6$He nucleus at $\theta=28^\circ$.
The eigenvalues of the $^4{\rm He}+n+n$  three-body and
$^5{\rm He}(3/2^-)+n$ two-body continuum states
appear reasonably along the lines which are
rotated from the real axis by $2\theta$.
We find two $2^+$ resonance poles at $E{\rm _r}=0.96$ MeV
with $\Gamma=0.14$ MeV and $E_r=2.81$ MeV with $\Gamma=4.63$ MeV,
respectively; the latter pole is stable against the change of $\theta$
from $25^\circ$ to $33^\circ$.  In addition, we find the $1^+$ state
to be at $E{\rm _r}=3.00$ MeV with $\Gamma=5.22$ MeV.
In Fig.~\ref{fig:he6level}, we summarize the energy spectra of $^6$He 
together with experimental data confirmed so far.
The calculated energies of the $0^+$ ground state and the first
$2^+$ excited state are in good agreement with the data.
It is interesting to obtain the second $2^+$ state at 2.81 MeV
above the $\alpha +n+n$ threshold with $\Gamma=4.63$ MeV.
Recently, the SPIRAL experiment suggests the existence of
the $2^+_2$ state at 1.63 MeV above $\alpha +n+n$ threshold
with $\Gamma =1.6$ MeV~\cite{Mougeot}. For comparison, here
we note that the $2^+_2$ state is calculated to be $\sim 1$ MeV
higher energy with a much broader width.

The calculation for the second $2^+$ state of $^6$He
has been already performed in Ref.~\cite{Myo2011} within the
cluster-orbital shell model (COSM) combined with the CSM.
They predicted this state to be at $E_r=2.52$ MeV with
$\Gamma=3.87$ MeV using the Minnesota $NN$ interaction.
Although the AV8 $NN$ interaction is employed in the present work,
we remark that both theoretical calculations give essentially
the same results with respect to the resonance energy and the
decay width.

Now we discuss the results for the hypernucleus $^7_{\Lambda}$He.
First, the calculated energy of the ground state ($J^\pi=1/2^+$) is
$E=-6.39$ MeV which corresponds to a $\Lambda$-separation energy
of $B_{\Lambda}=E(^6{\rm He})-E(^7_{\Lambda}{\rm He})=5.36$ MeV.
This energy is the same as in Ref.~\cite{Hiyama2009} in the case
of no CSB component in the $\Lambda N$ interaction.
The JLab E01-011 experiment measured the
$^7{\rm Li} (e,e'K^+) ^7_{\Lambda}$He reaction spectrum and
deduced~\cite{JLAB}  for the first time the $\Lambda$-separation
energy of the $^7_{\Lambda}$He ground state to be
$B_{\Lambda}^{\rm exp}=5.68 \pm 0.03 ({\rm stat.}) \pm
0.25({\rm sys.})$  MeV. The theoretical prediction of $B_\Lambda$
is compatible with this value within the experimental error.

Second, in the present calculation, the first $3/2^+$ and $5/2^+$
states are obtained as bound states at $E=-4.73$ MeV
and $E=-4.65$ MeV, respectively, with respect to the
$\alpha +n+n+\Lambda$ four-body breakup threshold; these values
are the same as in Ref.~\cite{Hiyama2009} in the case of
no CSB component in the $\Lambda N$ interaction.
It is noted that a recent analysis of the JLab E05-115
experiment~\cite{Gogami} also reports an excited-state peak at
$B_\Lambda^{\rm exp}=3.65 \pm 0.20 ({\rm stat.}) \pm 0.11({\rm sys.})$
MeV. This peak can be well attributed to the present prediction
of the $3/2^+_1$ and $5/2^+_1$ states, because the average
theoretical $\Lambda$-separation energy for these states,
$B_\Lambda=3.66$ MeV, is in very good agreement with the
experimental value. The agreement is confirmed also by the
comparison of production cross sections for the ground
and the excited states, since they are consistent with the
theoretical estimation as will be shown below.

\begin{figure}[t]
\begin{center}
\epsfig{file=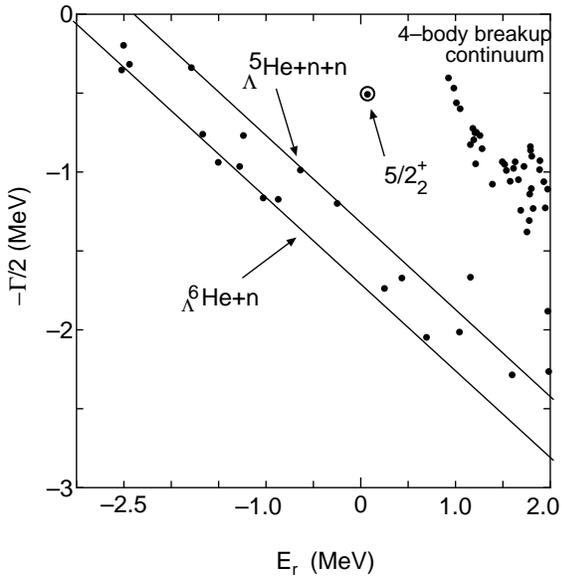,scale=0.45}
\end{center}
\caption{The $5/2^+_2$  eigenvalue distributions of the
complex scaled Hamiltonian with $\theta=15^\circ$.
The energy is measured with respect to $\alpha +\Lambda+n+n$
breakup threshold. Three solid lines are
$^6_{\Lambda}$He$+n$, $^5_{\Lambda}{\rm He}+n+n$ and
$^5$He$(3/2^-)+n+\Lambda$ threshold.
}
\label{fig:he7level}
\end{figure}

\begin{figure}[t]
\begin{center}
\epsfig{file=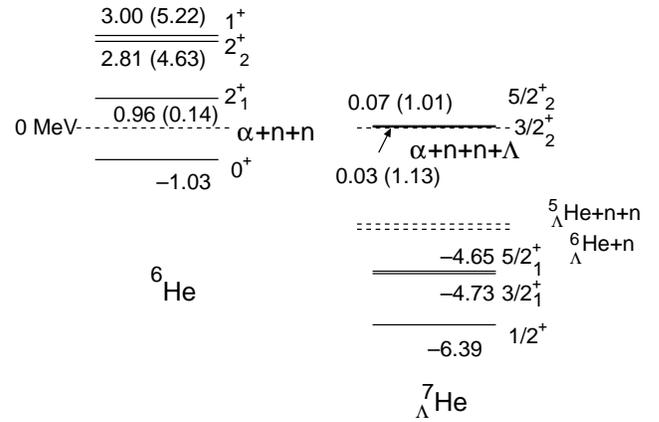,scale=0.41}
\end{center}
\caption{The calculated energy levels of
$^6$He and $^7_{\Lambda}$He.
The level energies in MeV are measured with respect to
$\alpha +n+n$ and $\alpha +\Lambda +n+n$
breakup thresholds.
The values in parentheses are decay widths $\Gamma$ in MeV.
}
\label{fig:he67level}
\end{figure}

Third, we search for the pole positions of the second
$3/2^+$ and $5/2^+$ state of $^7_{\Lambda}$He
within the complex scaling method (CSM). In Fig.~\ref{fig:he7level}, 
we show the distribution of complex eigenvalues of
the $5/2^+_2$ states at $\theta=15^\circ$
where the energy $E_r$ is measured from the
$\alpha +\Lambda +n+n$ four-body breakup threshold.
We find the resonance pole of the $5/2^+_2$ state at $E_{\rm r}=0.07$
MeV with $\Gamma/2= 0.51$ MeV. This state is isolated from the continuum
states of the $^6_{\Lambda}{\rm He}+n$ and $^5_{\Lambda}{\rm He}+n+n$
configurations and is stable against the change of the
rotation angle, $\theta= 10^\circ$ - $18^\circ$.
For the $3/2^+_2$ state, we find the resonance pole  at
$E_{\rm r}=0.03$ with $\Gamma/2=0.56$ MeV.
Figure~\ref{fig:he67level} summarizes the theoretical level structure of
$^7_{\Lambda}$He together with that of $^6$He.
In Fig.~\ref{fig:he67level}, we see a small energy splitting for the second 
$3/2^+$-$5/2^+$ doublet states which is similar to the splitting of
the first $3/2^+$-$5/2^+$ doublet. This is caused basically by the small 
spin-spin $\Lambda N$ interaction.  The $1^+$ state is obtained at 
$E_{\rm r}=3.00$ MeV with a decay  width $\Gamma=5.22$ MeV, very
close to that of the $2^+_2$ level, as shown in Fig.\ref{fig:he67level}.
However, we did not find any particular influence
of the $1^+$ state on the $3/2^+_2$ energy, probably because they 
have both large decay widths. One naturally expect to have  the third 
$3/2^+$ state and the second $1/2^+$ state in this region of excitation.
However, it was difficult to distinguish these states since these states 
were embedded into the four-body breakup continuum states.

Here we emphasize that, when the $\Lambda$ particle is added, the
responses of two $2^+$ core states are so different that
the energy spacing between the centroids of the first doublet
($3/2_1^+$, $5/2_1^+$) and the second doublet ($3/2_2^+$, $5/2_2^+$)
in $^7_{\Lambda}$He becomes quite large ($\sim 4.7$ MeV) in
comparison with the 1.85 MeV spacing of the two $2^+$ states.
This effect is attributed to the difference in size (spatial structure)
of the two $2^+$ wave functions in $^6$He. In fact the
decay width of the $2_1^+$ resonant state is very small (0.11 MeV),
and the state is compact.
On the other hand, the decay width of the $2_2^+$  state is
considerably large (4.63 MeV). Therefore it is expected that the radial extent
of the wave function for the $2_2^+$ state is much larger than that
of the $2_1^+$ state.
When a $\Lambda$ particle is added to such states having different
characters, the energy gain, $\Lambda$ separation energy, in the
compact state is much larger than that in the dilute state.
A similar phenomenon has been pointed out in Ref.~\cite{Hiyama2000},
namely that the $\Lambda$ separation energy of the compact shell-like
ground state, $1/2^+_1$ in $^{13}_{\Lambda}$C, is much larger
than that in a well-developed clustering state such as the $1/2^+_2$ state.
If the predicted second doublet states ($3/2_2^+$ and $5/2_2^+$) is
confirmed in a future experiment, then we can see the clear
state-dependent response to the addition of a $\Lambda$ particle.


Next, it is interesting to investigate
how we can confirm experimentally the energy positions of
the $3/2^+_2$ and $5/2^+_2$ states predicted here in
$^7_{\Lambda}$He. In view of the successful E05-115
experiment~\cite{JLAB,Gogami} which already shows the ground
state and the excited state peaks, the unique possibility is to
perform a dedicated $^7{\rm Li}(e,e'K^+)^7_{\Lambda}$He
reaction experiment with better energy resolution in the future.
For the sake of experimental feasibility study, here we give
brief estimates of the $^7{\rm Li} (\gamma,K^+) ^7_{\Lambda}$He
reaction cross sections in DWIA on the basis of the COSM
results of the proton pickup
spectroscopic amplitudes $S^{1/2}$ which are ready to be
used.  As for the $^7$Li target state with strong binding, 
we use the ordinary shell-model wave function of the maximum 
symmetric \mbox{ $|p^3(30)_{L=1}(T=1/2,S=1/2);J=3/2^->$} which corresponds 
to the $\alpha-t$ cluster configuration in the lowest approximation. 
However,  one should be careful in treating  weakly bound or 
unbound states in $^6$He. 

The low-lying states of
$^6$He have been studied extensively by the COSM with $\alpha$
core in Ref.~\cite{Myo2011} which treats the many-body
resonances in the complex scaling method (CSM).
Here we remark that the present three-body cluster model and
the COSM treatment give essentially the same physical
properties of nuclei. As explained in
Refs.~\cite{Ikeda92,Aoyama2001}, the choice of the
relative coordinates in the model space of core+$n$+$n$
is different but the COSM gives the consistent results of
core$+n+n$ three-cluster model for the structures such
as $^6$He and $^{11}$Li~\cite{Myo2011}.
It is also confirmed that the COSM and Gamow shell model
(GSM) give the same results for the $^6$He structure for
the energy eigenvalues, configuration mixing, and the
density distribution of halo neutrons~\cite{myo2014,Masui2014}.
Instead of using the 4-body cluster model wave functions, we
make use of the proton pickup spectroscopic amplitudes
$S^{1/2}$ derived in the COSM framework, and correspondingly
we assume the simplified weak-coupling wave functions consisting
of the $^6$He($0^+, 2^+_1, 2^+_2$) solutions
and $s_{1/2}^{\Lambda}$ in the brief estimates of the cross sections.

One may refer to Table II of Ref.~\cite{Myo2011} for the
$^6$He($0^+$) bound state wave function, while the two $2^+$
resonance states of $^6$He have the following structures,
respectively, showing the dominant components  symbolically.
\begin{eqnarray}
 (2^+_1)= \sqrt{0.898+i0.013}[p_{3/2}^2]
                +\sqrt{0.089-i0.013}[p_{3/2}p_{1/2}], \nonumber\\
  (2^+_2) =\sqrt{0.089-i0.023}[p_{3/2}^2]
               -\sqrt{0.889+i0.024}[p_{3/2}p_{1/2}]. \nonumber
\end{eqnarray}

Here the amplitude of each COSM component is complex 
reflecting the calculated decay width.  It is also notable that
these wave functions are much different from those of the
usual SU(3)($\lambda\mu$) prescription like
$|p^2(20)_{L=2}$($T$=1,$S$=0);$J=2^+_1>$ and
$|p^2(01)_{L=1}$($T$=1,$S$=1);$J=2^+_2>$,
because the $(p_{3/2}^2)$ component is smaller than
$(p_{3/2}p_{1/2})$ in $J=2_1^+$ in this prescription;
the two components are almost equally admixed
in conventional shell-model calculations~\cite{Kurath}.
The essential reason for the difference is attributed
to the fact that in the COSM treatment the experimental
spin-orbit splitting for the unbound $p$-state neutron
on $\alpha$ is properly taken into account through a
realistic $\alpha N$ potential~\cite{KanadaKKNN,Aoyama2001}
consistent with the phase shift analysis.
It should be also noted that in the present three- and four-body calculations,
the same $\alpha N$ potential is employed.
For readers' reference, we remark that in COSM
the $(p_{3/2}p_{1/2})_{J=2}$ single-channel energy
is estimated to be 1.28 MeV higher
than the $(p_{3/2}^2)_{J=2}$ channel. It is also noted that
the proton-pickup spectroscopic factors from
the $^7$Li($3/2^-_{gs}$) leading to the $^6$He states
are calculated to be $C^2S=$ 0.559 ($0^+$),
0.257 ($2^+_1$), and 0.097 ($2^+_2$), by neglecting
the small imaginary components. If one uses the
conventional shell-model wave functions with an unrealistically
small spin-orbit splitting, the $S$-factor leading to the $2^+_2$
state is almost vanishing.

The $^7{\rm Li} (\gamma,K^+) ^7_{\Lambda}$He differential cross
sections are estimated at $E_{\gamma}^{Lab}$=1.5 GeV and
$\theta_K^{Lab}$=7 deg corresponding  the E05-115 experimental
kinematics.
The calculated results are:
\begin{eqnarray}
 d\sigma/d\Omega(1/2^+_G)&=& 49.0  {\rm nb/sr} \label{eq:3} \\
 d\sigma/d\Omega(3/2^+_1+5/2^+_1)&=&10.0+11.6=21.6 {\rm nb/sr}, \label{eq:4}\\
 d\sigma/d\Omega(3/2^+_2+5/2^+_2)&=&3.4+4.3=7.7 {\rm nb/sr}.\label{eq:5}
\end{eqnarray}
Here the two doublet strengths are summed up, as they are degenerate
in energies. If one chooses
$\theta_K^{Lab}$=2 deg, then the differential cross sections
corresponding to Eqs.(\ref{eq:3})-(\ref{eq:5}) are
estimated to be 73.8 nb/sr, 32.7 nb/sr, and 11.6 nb/sr, respectively.
We remark that the relative strength for the ground state and the
degenerate excited states (49.0nb/sr vs. 21.6 nb/sr) is in very good
agreement with the ground and second peaks observed in the
JLab E05-115 experiment~\cite{JLAB,Gogami}.
Thus we can surely expect that the $3/2^+_2$ and $5/2^+_2$ states
should appear as the third narrow peak having a differential
cross section of about 40\% of that for the second peak.
If this peak position is identified in a future experiment,
it will open an important window toward the study of
characteristic structures of neutron-rich hypernuclei.

As for the excited states of $^6$He, we note that the energy position
of the second $2^+_2$ state is calculated to be $\sim 1$ MeV higher,
with much broader width, than the SPIRAL experimental
data~\cite{Mougeot} that provides the $2^+_2$ state energy for
the first time. In view of the theory vs. experiment discrepancies in
energy and width for this $2^+_2$ resonance state, we suggest
new experiments to search for this state in $^6$He for reconfirmation.
We emphasize that such experiments to study the typical neutron-halo 
nuclear excited states, combined with a dedicated
$^7{\rm Li} (e,e'K^+) ^7_{\Lambda}$He experiment, could provide
a new possibility for the spectroscopy of neutron-rich hypernuclei.

\section{Summary}

Motivated by the recent data of $^7_{\Lambda}$He at JLab,
we calculated resonant states of $3/2^+_2$ and   $5/2^+_2$ within the
framework of $\alpha +n+n+\Lambda$ four-body cluster model using CSM.
The resonant $2^+_2$ state of the core nucleus $^6$He was
calculated within the framework of
$\alpha +n+n$ three-body cluster model.
All the two-body interactions
among subunits for $^6$He and $^7_{\Lambda}$He are the same as
in Ref.~\cite{Hiyama2009}, that is, the interactions are  chosen
to reproduce the binding  energies of all subsystems composed of
two and three subunits.
Then, we obtained that the energy of second $2^+$ state was
2.81 MeV with $\Gamma=4.63$ MeV for $^6$He.
This calculated energy is located above by $\sim 1$ MeV
and narrower than the recent data from SPIRAL~\cite{Mougeot}.
For $^7_{\Lambda}$He, we predicted the energies of the second
$3/2^+$ and $5/2^+$ resonant states to be 0.03 and 0.07 MeV
with $\Gamma=\sim 1 $ MeV, which have narrower widths than
the corresponding state of $^6$He core nucleus, $2^+_2$,
due to the gluelike role of the $\Lambda$ particle.

To encourage future experiments at JLab, we present brief estimates
for the $^7{\rm Li} (\gamma, K^+) ^7_{\Lambda}$He reaction cross
sections by making use of the spectroscopic amplitudes derived
from the COSM framework. The calculated cross sections are 3.4 nb/sr
and 4.3 nb/sr for $3/2^+_2$ and $5/2^+_2$, respectively, if the E05-115
experimental kinematics is assumed. Thus we predict that
the first doublet peak ($3/2^+_1$ and $5/2^+_1$) should have
the strength of about $40$ \% of the ground state peak, while
the second doublet should form the third peak having about
$15$ \% which seems a feasible strength for a future experiment.
For further reconfirmation, the wave functions within the
four-body cluster model framework for $\alpha +n+n+p$ and
$\alpha +\Lambda +n+n$ will be also applied to the cross section
estimates and the results will be discussed in the forthcoming
paper. 

In conclusion we have shown the importance of measuring
the third peak consisting of the $3/2_2^+$ and $5/2_2^+$
resonant states of $^7_{\Lambda}$He together with the
experimental confirmation of the second $2^+$ state energy
and width of the core nucleus $^6$He.
The predicted large changes of binding energies and widths
before and after the $\Lambda$ addition are interesting
and rich aspects to be realized in neutron-rich hypernuclei.
Thus, we suggest that measurements should be performed to confirm the
second $2^+$ resonant state of $^6$He on one hand and, on the
other hand, to find the second $3/2^+$ and $5/2^+$ doublet
states in a dedicated $^7{\rm Li}(e,e'K^+) ^7_{\Lambda}$He
experiment.

\section*{Acknowledgments}
The authors thank Professors S.N. Nakamura, S. Kubono. T. Nakamura,
B. F. Gibson, H. Ohtsu and Dr. Gogami
for informative discussions.
This work was supported by a Grant-in-Aid for Scientific Research from
Monbukagakusho of Japan.  The numerical calculations were performed
on the HITACHI SR16000 at KEK and YITP. This work was partly
supported by JSPS Grant No. 23224006 and by RIKEN iTHES Project.


\begin{thebibliography}{99}

\bibitem{JLAB} S. N. Nakamura {\it et al.}, Phys.\ Rev.\ Lett.\
{\bf 110}, 012502 (2013).

\bibitem{Hiyama96} E. Hiyama, M. Kamimura, T. Motoba,
T. Yamada, Y. Yamamoto, Phys. Rev. C{\bf 53},  2075 (1996).

\bibitem{Hiyama2009} E. Hiyama, Y. Yamamoto, and
M. Kamimura, Phys. Rev. C{\bf 80}, 054321 (2009).

\bibitem{Gogami} T. Gogami, PhD Thesis, Tohoku University, 2014.

\bibitem{Tilley} D. R. Tilley {\it et al.}, Nucl Phys. A {\bf 708}, 3 (2002).

\bibitem{Nakamura2000} T. Nakamura {\it et al.},
 Phys. Lett. B {\bf 493}, 209 (2000).

\bibitem{Mougeot} X. Mougeot {\it et al.}, Phys. Lett. B {\bf 718}, 441 (2012).

\bibitem{Pieper} S. C. Pieper, R. B. Wiringa, and J. Carlson,
Phys. Rev. C {\bf 70}, 054325 (2004).

\bibitem{Hagen} G. Hagen, M. Hjortj-Jensen, and J. S. Vaagen,
Phys. Rev. {\bf 71}, 044314 (2005).

\bibitem{Volya} A. Volya, and  V. Zelevinsky,
 Phys. Rev. Lett. {94}, 052501 (2005).

\bibitem{Myo2011} T. Myo, K. Kato and K. Ikeda, Phys. Rev. C {\bf 76},
054309 (2007); T. Myo, R. Ando and K. Kato, Phys. Rev. C {\bf 80}, 014315
(2009);T. Myo, Y. Kikuchi, and K. Kato, Phys. Rev. C {\bf 84},
064306 (2011).

\bibitem{Hiyama2003} E. Hiyama, Y. Kino, and M. Kamimura,
Prog.\ Theor.\ Nucl.\ Phys.\ {51}, 223 (2003).

\bibitem{CSM-ref1}     J.~Aguilar, and J.M.~Combes,
      Commun. Math. Phys. \textbf{ 22}, 269 (1971).

\bibitem{CSM-ref2}     E.~Balslev, and J.M.~Combes,
      Commun. Math. Phys. \textbf{ 22}, 280 (1971).

\bibitem{CSM-ref3}     B.~Simon,
      Commun. Math. Phys. \textbf{ 27}, 1 (1972).

\bibitem{Ho} Y. K. Ho, Phys. Rep. {\bf 99}, 1 (1983).

\bibitem{Moiseyev} N. Moiseyev, Phys. Rep. {\bf 302}, 211 (1998).

\bibitem{Aoyama2006} S. Aoyama, T. Myo, K. Kato,
and K. Ikeda, Prog. Theor. Phys. {\bf 116}, 1 (2006).

\bibitem{myo2014}
T. Myo, Y. Kikuchi, H. Masui, and K. Kato, Prog. Part.
 Nucl. Phys. {\bf 79}, 1 (2014).

\bibitem{Hiyama2000}  E. Hiyama, M. Kamimura, T, Motoba,
T. Yamada, and Y. Yamamoto, Phys. Rev. Lett. 85, 270 (2000).

\bibitem{Ikeda92} K. Ikeda, Nuclear Physics A538, 355c  (1992).

\bibitem{Aoyama2001}  S. Aoyama, K. Kato and K. Ikeda, Prog. Theor.
 Phys. Supple. 142  35 (2001).

\bibitem{Masui2014}  H. Masui,  K.Kato,  N.Michel, and
 M.Ploszajczak, Phys. Rev.  C {\bf 89} 044317  (2014).

\bibitem{Kurath} S. Cohen and D. Kurath, Nucl. Phys.
{\bf 73}, 1 (1965).

\bibitem{KanadaKKNN} H. Kanada, T. Kaneko, S. Nagata, and M. Nomote,
  Prog. Theor. Phys. {\bf 61}, 1327 (1975).

\end{thebibliography}
\end{document}